\documentclass[twocolumn,showpacs,superscriptaddress,aps,pra]{revtex4-1}
\usepackage{amssymb}
\usepackage{bm}
\usepackage{graphicx}
\usepackage{dcolumn}
\usepackage{amsmath,txfonts}
\usepackage{epstopdf}
\usepackage[mathlines]{lineno}
\usepackage{color}
\usepackage[colorlinks=true,linkcolor=blue,citecolor=blue]{hyperref}
\usepackage[normalem]{ulem}

\begin{document}

\title{Nonreciprocal quantum phase transition in cavity magnonics}
\author{Ye-Jun Xu}
\email{yejunxu@126.com}
\author{Long-Hua Zhai}
\author{Peng Fu}
\author{Shou-Jing Cheng}
\affiliation{Interdisciplinary Research Center of Quantum and
	Photoelectric Information, and Anhui Research Center of Semiconductor Industry Generic Technology, Chizhou University, Chizhou, Anhui, 247000, China}

\author{Guo-Qiang Zhang}
\email{zhangguoqiang@hznu.edu.cn}

\affiliation{School of Physics, Hangzhou Normal University, Hangzhou
	311121,	China}

\begin{abstract}
We investigate the nonreciprocal quantum phase transition in a cavity
magnonic system driven by a parametric field, where an yttrium iron garnet
(YIG) sphere is placed in a spinning microwave resonator. The system
exhibits a rich phase diagram due to both magnon Kerr nonlinearity in YIG
and parametric drive on the resonator. Especially, Sagnac-Fizeau shift
caused by the spinning of the resonator brings about a significant
modification in the critical driving strengths for second- and first-order
quantum phase transitions, which means that the highly controllable quantum
phase can be realized by the spinning speed of the resonator. More
importantly, based on the difference in the detunings of the
counterclockwise and clockwise modes induced by spinning direction of the
resonator, we show that the phase transition in this system is
nonreciprocal, that is, the quantum phase transition occurs when the cavity
is driven in one direction but not the other. Our work offers an alternative
path to engineer and design nonreciprocal magnonic devices.
\end{abstract}

\date{\today }
\maketitle

\section{Introduction}

In recent years, the cavity magnonic system has become an outstanding
platform for exploring light-matter interactions~\cite%
{Lachance-Quirion2019,Rameshti2022,PengYan-2021}. Since ferrimagnetic
systems, e.g., yttrium iron garnet (YIG), own high spin density and low
damping rate, the strong (even ultra-strong) coupling between magnons in
ferrimagnetic systems and photons in microwave cavities can be achieved
experimentally~\cite%
{Huebl2013,Tabuchi2014,Zhang2014,Goryachev2014,Hu-PRL-15,Zhang-npjQI-15}.
This strong coherent interaction allows one to study many fascinating
phenomena, such as magnon dark modes \cite{01}, magnon-photon entanglement
\cite{01a}, magnon blockade \cite{01b,01c,01d}, non-Hermitian physics \cite%
{02,01e,Zhang19,01f,01g}, quantum states of magnons \cite%
{01h,01i,01j,Zhang23}, cooperative polariton dynamics \cite{01k}, and magnon
spintronics \cite{01n,03}. In addition, it is worth mentioning that the
magnon Kerr effect, originating from the magnetocrystalline anisotropy in
the YIG, has been theoretically and experimentally demonstrated in
cavity-magnon systems \cite{03a,03b}. The magnon Kerr effect gives rise to
the development of some topics, including bistability and tristability of
cavity-magnon polaritons~\cite{04,Nair20,Bi21,Shen21,Bi24}, high-order
sideband generation \cite{05}, coherent perfect absorption~\cite%
{Zhang-Wang23}, quantum entanglement \cite{Yang21,Zhang-Agarwal19}, and
long-range spin-spin coupling \cite{08}.

Owing to its crucial role in quantum physics and potential applications in
quantum technologies, the quantum phase transition (QPT) has attracted
considerable attention\cite{1,1a,2,2a,2b,3,4,5,6,7,9,10,11,12,13,14}.
Different from the classical phase transition happening at a finite
temperature, the QPT mainly originates from the quantum fluctuation. In the
critical parameter regime, the QPT occurs between two stable phases
accompanied by spontaneous symmetry breaking as the temperature tends to the
absolute zero of temperature. Recently, QPT has not only been investigated
in the Dicke model (or the Tavis-Cummings model) without the ultrastrongly
coupling requirement when the cavity field is squeezed via parametric drive~%
\cite{18}, but also been studied in dissipative systems with other types of
nonlinearity (e.g., Kerr nonlinearity) \cite{19,20}. With the assistance of
magnon Kerr effect and parametric drive, the QPT has also been explored in
cavity magnonics \cite{21,22}.

Nonreciprocal physics refers to the phenomenon that a system displays
different responses in opposite directions. Up to now, the study of
nonreciprocity has been extended to many scientific branches, such as optics
\cite{23,24}, acoustics \cite{25,26,27}, and thermodynamics \cite{28}.
Traditional nonreciprocal devices break Lorentz reciprocity to achieve
nonreciprocity mainly by employing the Faraday effect in magneto-optical
crystal materials \cite{29,30}. In such nonreciprocal devices, the used
magnetic materials are bulky due to the requirement of high susceptibility
to external magnetic field interference, which makes them unsuitable for
on-chip integration. To circumvent these obstacles, a variety of
nonreciprocity schemes have been proposed based on nonlinear optics \cite%
{31,32,32a,32b}, optomechanics \cite{33,34,35}, non-Hermitian optics \cite%
{36,37,38,39}, etc. In particular, a recent experiment confirmed that an
optical diode with 99.6\% isolation has been realized by using a spinning
resonator \cite{40}, in which the nonreciprocity is caused by the Fizeau
shift of circulating lights. Subsequently, there are increasing interests in
the connection of the nonreciprocity with other quantum effects, such as
quantum blockades \cite{41,42,42a,43,44,45,46,47,47a,48}, quantum
entanglement \cite{49,50,51,Chen23,52}, mechanical squeezing \cite{53,54,55}%
, phonon and magnon laser \cite{56,57,58}, sideband responses \cite{59,60},
single-photon state conversion \cite{60a}, and superradiant phase
transitions \cite{61c,61d,61}. However, the nonreciprocal QPT has not yet
been investigated in cavity magnonics.

Here we propose a scheme to realize a nonreciprocal QPT in a spinning
microwave magnonic system. We first display the phase diagram of the hybrid
system, which contains parity-symmetric phase (PSP), parity-symmetry-broken
phase (PSBP), and bistable phase (BP).\textrm{\ }In particular, the Sagnac
effect caused by the spinning of the resonator can result in different
influences on the detuning of the counterclockwise (CCW) and clockwise (CW)
fields compared to the driving field. This leads to the occurrence of the
nonreciprocal QPT for driving the resonator from the opposite directions. By
investigating the different behaviors of the mean magnon number in the
vicinity of the critical point for different Fizeau shifts, we show that the
QPT is not only either continuous (second-order) or discontinuous
(first-order), but also nonreciprocal for different driving directions.
Moreover, an isolation parameter is introduced to quantitatively describe
the nonreciprocal QPT. At last the nonreciprocal fluctuation of magnon
number is also numerically simulated for displaying the features of the
nonreciprocal QPT from the point of view of quantum fluctuation.
Nonreciprocal QPT may open up the prospect of engineering nonreciprocal
devices for applications in, e.g., on-chip unidirectional quantum sensing
\cite{61a} and quantum metrology \cite{61b}.

In other systems, nonreciprocal phase transition has been investigated~\cite{61c,61d,61}.
Especially, Ref.~\cite{61c} makes a significant
contribution to the theory of nonreciprocal phase transition, which originates from
asymmetric interactions of multiple species. Subsequently,
nonreciprocal coupling is further used to realize nonreciprocal QPT in an open Dicke model~\cite{61d}.
Very recently, Zhu {\it et al}~\cite{61} proposed to study nonreciprocal QPT
in an open dual-coupling Jaynes-Cummings model by utilizing Sagnac effect, where a
two-level atom is nonlinearly coupled to modes of a spinning microcavity.
As an advantage of this work, it is
demonstrated that in the presence of cavity dissipation, the squeezed light can recover the QPT
through breaking the requirement of ultrastrong atom-field coupling.
On the contrary, although we propose to achieve the
nonreciprocity of QPT based on Sagnac effect, our scheme is
established in a spinning microwave magnonical system. Here the achievement
of QPT relies on not only the squeezed field, but also the magnon Kerr
nonlinearity in the YIG. Our scheme exploits a different mechanism to
engineer nonreciprocal QPT.

The remainder of this paper is organized as follows. In Sec. II, we
introduce the theoretical model. Using the Heisenberg-Langevin approach, we
give the steady-state solutions of the system. In Sec. III, the
nonreciprocal QPT, induced by the Sagnac-Fizeau shift, is studied in detail.
Finally, we summarize our results in Sec. IV.

\section{The model}

\begin{figure}[h]
\centering\includegraphics[width=8cm]{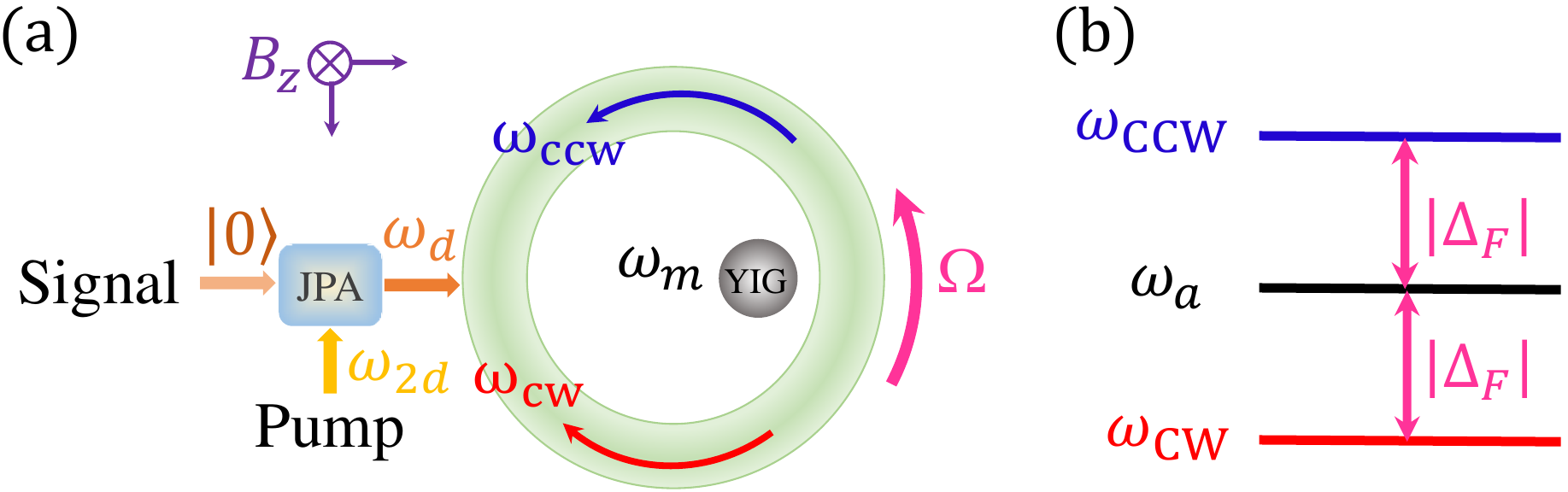}
\caption{(a) Schematic of the system. A YIG sphere, biased by a uniform
magnetic field $B_z$ along the $z$ direction, is placed inside a spinning
microwave resonator driven by a squeezed light generated by a flux-driven
JPA. (b) Frequencies of the microwave resonator. When the microwave
resonator rotates in the fixed counterclockwise (CCW) direction, the
resonance frequencies of the CCW mode and clockwise (CW) mode will
experience different Fizeau shifts, where the frequencies $\protect\omega %
_{a}+\left\vert \Delta _{F}\right\vert $ and $\protect\omega _{a}-\left\vert
\Delta _{F}\right\vert $ denote for the CCW and CW modes, respectively.}
\label{fig1}
\end{figure}

As schematically shown in Fig. \ref{fig1}, we consider a cavity-magnon
system consisting of a spinning resonator and a YIG sphere, where photons in
the resonator are coupled to magnons in YIG. In the hybrid system, the
resonator is driven by a weak squeezed vacuum field generated by a
flux-driven Josephson parametric amplifier (JPA) and the magnetocrystalline
anisotropy of YIG results in the Kerr nonlinear interaction among magnons.
The Hamiltonian of the total system reads ($\hbar =1$)%
\begin{eqnarray}
H &=&\left( \omega _{a}+\Delta _{F}\right) a^{\dagger }a+\omega
_{m}m^{\dagger }m+\frac{K}{2}m^{\dagger }m^{\dagger }mm  \notag \\
&&+J\left( a^{\dagger }m+am^{\dagger }\right) +\frac{G}{2}\left( a^{\dagger
2}e^{-2i\omega _{d}t}+a^{2}e^{2i\omega _{d}t}\right) ,
\end{eqnarray}%
where $a$ $\left( a^{\dagger }\right) $ and $m$ $\left( m^{\dagger }\right) $
are, respectively, the annihilation (creation) operators of the cavity and
magnon modes with the corresponding resonance frequencies $\omega _{a}$ and $%
\omega _{m}$. The microwave resonator is parametrically driven with driving
amplitude $G$ and pump frequency $\omega _{d}$. $K$ denotes the nonlinear
coefficient of the magnon Kerr effect and $J$ is the strength of the
cavity-magnon interaction. For a resonator spinning at an angular velocity $%
\Omega $, the frequencies of the CCW and CW modes experience Sagnac-Fizeau
shift, i.e., $\omega _{a}\rightarrow \omega _{a}+\Delta _{F}$, with \cite%
{40,62}%
\begin{equation}
\Delta _{F}=\pm \Omega \frac{nr\omega _{a}}{c}\left( 1-\frac{1}{n^{2}}-\frac{%
\lambda }{n}\frac{dn}{d\lambda }\right) ,
\end{equation}%
in which $n$ is the refractive index, $r$ is the radius of the resonator,
and $\lambda $ $(c)$ is the wavelength (speed) of the light in vacuum. The
dispersion term $dn/d\lambda $, characterizing the relativistic origin of
the Sagnac effect, is relatively small ($\sim $1\%) and thus can be ignored.

In the rotating frame with respect to the driven frequency $\omega _{d}$,
the Hamiltonian of the system becomes
\begin{eqnarray}
\mathcal{H} &=&\left( \tilde{\Delta}_{a}-i\kappa \right) a^{\dagger }a
     +\left( \Delta_{m}-i\gamma \right) m^{\dagger }m
     +\frac{K}{2}m^{\dagger }m^{\dagger }mm \notag \\
  & &+J\left( a^{\dagger }m+am^{\dagger }\right)
     +\frac{G}{2}\left( a^{\dagger2}+a^{2}\right) ,  \label{h1}
\end{eqnarray}%
where the dissipations of the cavity-magnon system have been considered.
Here $\kappa $ ($\gamma $) is the damping rate of the photon (magnon) mode,
$\tilde{\Delta}_{a}=\Delta _{a}+\Delta _{F}$, $\Delta _{a}=\omega
_{a}-\omega _{d}/2$, and $\Delta _{m}=\omega _{m}-\omega _{d}/2$. Note that
in the absence of YIG sphere (i.e., $J=0$), the squeezing driving can cause
the system unstable if $G>\sqrt{\tilde{\Delta}_{a}^{2}+\kappa ^{2}}$ \cite%
{63}. Hence we only consider the case of $G<\sqrt{\tilde{\Delta}%
_{a}^{2}+\kappa ^{2}}$.
Interestingly, the non-Hermitian Hamiltonian $\mathcal{H}$ in Eq.~(\ref{h1}) commutes
with the parity operator $\mathcal{P} =\exp \left[ i\pi \left(
a^{\dagger }a+m^{\dagger }m\right) \right] $\ \cite{2}, i.e., $\left[
\mathcal{H},\mathcal{P} \right] =0$, which indicates that the Hamiltonian $\mathcal{H}$ has
 the parity symmetry. Normally, the total excitation number of the
system at steady states can be any natural number. When it takes odd (even) numbers, the
steady state of the system has the odd (even) parity. If the total excitation number
of the system versus system parameters varies in the even subset $\{0, 2, 4, 6, 8,\cdots\}$
or the even subset $\{1, 3, 5, 7, 9,\cdots\}$, we say that the corresponding steady state of the system is
parity-symmetric. Otherwise, the steady state of the system is parity-symmetry-breaking.

By considering dissipation and noise effects in the Hamiltonian (\ref{h1})
\cite{64}, the quantum Langevin equations (QLEs) describing the dynamics of
the hybrid system can be written as
\begin{eqnarray}
\dot{a} &=&-i\left( \tilde{\Delta}_{a}-i\kappa \right) a-iJm-iGa^{\dagger }+%
\sqrt{2\kappa }a_{\text{in}},  \notag \\
\dot{m} &=&-i\left( \Delta _{m}-i\gamma \right) m-iKm^{\dagger }mm-iJa+\sqrt{%
2\gamma }m_{\text{in}},  \label{5}
\end{eqnarray}%
where $a_{\text{in}}$ and $m_{\text{in}}$ are the zero-mean input noise
operator for the microwave photon (magnon) mode, respectively. To linearize
the Eq. (\ref{5}), we can expand the operators $a$ and $m$ as a summation of
their expectation values and quantum fluctuations, i.e., $a=A+\delta a$ and $%
m=M+\delta m$. Following from Eq. (\ref{5}), the dynamical equations for the
expectation values $A$ and $M$ are
\begin{eqnarray}
\dot{A} &=&-i\left( \tilde{\Delta}_{a}-i\kappa \right) A-iJM-iGA^{\ast },
\notag \\
\dot{M} &=&-i\left( \Delta _{m}-i\gamma \right) M-iK\left\vert M\right\vert
^{2}M-iJA.  \label{6}
\end{eqnarray}%
In the steady-state case ($\dot{A}=\dot{M}=0$), solving Eq. (\ref{6}) gets
three solutions for the mean magnon number $\left\vert M\right\vert ^{2}$,
i.e.,
\begin{equation}
\left\vert M\right\vert _{0}^{2}=0,\text{ }\left\vert M\right\vert _{\pm
}^{2}=\frac{-\tilde{\Delta}_{m}\pm \sqrt{\beta ^{2}G^{2}-\tilde{\gamma}^{2}}%
}{K},  \label{6d}
\end{equation}%
with $\tilde{\Delta}_{m}=\Delta _{m}-\beta \tilde{\Delta}_{a}$, $\tilde{%
\gamma}=\gamma +\beta \kappa $, and $\beta =J^{2}/$($\tilde{\Delta}%
_{a}^{2}+\kappa ^{2}-G^{2}$). According to the first equation in Eq. (\ref{6}%
), the steady-state photon occupation can be expressed as%
\begin{equation}
\left\vert A\right\vert _{0}^{2}=0,\text{ }\left\vert A\right\vert _{\pm
}^{2}=\frac{\left( \tilde{\Delta}_{m}+\left\vert M\right\vert _{\pm
}^{2}\right) ^{2}+\gamma ^{2}}{g^{2}}\left\vert M\right\vert _{\pm }^{2}.
\end{equation}%
Thus, we need only to consider the mean magnon number $\left\vert
M\right\vert ^{2}$ as an order parameter to characterize the QPT of the
system. In order to derive critical conditions of the phase transition, we
first concentrate our attention on $\left\vert M\right\vert _{+}^{2}\geq 0$.
In the case of $-\tilde{\Delta}_{m}\geq 0$ (i.e., $\Delta _{m}/\tilde{\Delta}%
_{a}\leq \beta $), we take $\beta ^{2}G^{2}-\tilde{\gamma}^{2}=0$ in Eq. (%
\ref{6d}), which gives rise to $G=G_{\text{c1}}$ with
\begin{equation}
G_{\text{c1}}=\frac{-g^{2}+\sqrt{4\gamma ^{2}\tilde{\Delta}_{a}^{2}+\left(
g^{2}+2\gamma \kappa \right) ^{2}}}{2\gamma }.  \label{6d1}
\end{equation}%
Thus the critical driving strength $G_{\text{c1}}$ is independent of the
detuning $\Delta _{m}$. In the other case, when $-\tilde{\Delta}_{m}\leq 0$
and $\sqrt{\beta ^{2}G^{2}-\tilde{\gamma}^{2}}\geq 0$, the critical driving
strength is
\begin{equation}
G_{\text{c2}}=\sqrt{\frac{\left( g^{2}-\tilde{\Delta}_{a}\Delta _{m}\right)
^{2}+\gamma ^{2}\tilde{\Delta}_{a}^{2}+2g^{2}\gamma \kappa +\gamma
^{2}\kappa ^{2}+\Delta _{m}^{2}\kappa ^{2}}{\gamma ^{2}+\Delta _{m}^{2}}}.
\label{6e}
\end{equation}%
It clearly see from Eqs. (\ref{6d1}) and (\ref{6e}) that both critical
driving strengths are subject to the Fizeau shift $\Delta _{F}$. Secondly,
we can use the nontrivial solution $\left\vert M\right\vert _{-}^{2}$ to
obtain the third phase transition point even if it is proved to be unstable
in the following. By only considering the case $-\tilde{\Delta}_{m}\geq 0$
and $\sqrt{\beta ^{2}G^{2}-\tilde{\gamma}^{2}}\geq 0$, we obtain the
critical driving strength $G=G_{\text{c2}}$, which has the same form as Eq. (%
\ref{6e}).

On the other hand, the linearized QLEs for quantum fluctuations can be
written as

\begin{eqnarray}
\delta \dot{a} &=&-i\left( \tilde{\Delta}_{a}-i\kappa \right) \delta
a-iJ\delta m-iG\delta a^{\dagger }+\sqrt{2\kappa }a_{\text{in}},  \notag \\
\delta \dot{m} &=&-i\left( \tilde{\Delta}_{m}-i\gamma \right) \delta
m-iJ\delta a-iKM^{2}\delta m^{\dagger }+\sqrt{2\gamma }m_{\text{in}},  \notag
\\
&&  \label{7}
\end{eqnarray}%
with $\tilde{\Delta}_{m}=\Delta _{m}+2K\left\vert M\right\vert ^{2}$. Here
we have neglected the high-order terms of the fluctuations. By defining the
quadrature operators $\delta Q=\left( \delta a^{\dagger }+\delta a\right) /%
\sqrt{2}$, $\delta P=i\left( \delta a^{\dagger }-\delta a\right) /\sqrt{2}$,
$\delta X=\left( \delta m^{\dagger }+\delta m\right) /\sqrt{2}$, and $\delta
Y=i\left( \delta m^{\dagger }-\delta m\right) /\sqrt{2}$, Eq. (\ref{7}) can
be rewritten in a compact matrix form%
\begin{equation}
\delta \mathbf{\dot{O}}=\mathbf{U}\cdot \delta \mathbf{O}+\mathbf{O}_{\text{%
in}},  \label{8a}
\end{equation}%
with the vector of quadrature components $\delta \mathbf{O=}\left( \delta
Q,\delta P,\delta X,\delta Y\right) ^{\text{T}}$ and the vector of noise
quadratures $\mathbf{O}_{\text{in}}=\left( \sqrt{2\kappa }Q_{\text{in}},%
\sqrt{2\kappa }P_{\text{in}},\sqrt{2\gamma }X_{\text{in}},\sqrt{2\gamma }Y_{%
\text{in}}\right) ^{\text{T}}$ with $Q_{\text{in}}=\left( a_{\text{in}%
}^{\dagger }+a_{\text{in}}\right) /\sqrt{2}$, $P_{\text{in}}=i\left( a_{%
\text{in}}^{\dagger }-a_{\text{in}}\right) /\sqrt{2}$, $X_{\text{in}}=\left(
m_{\text{in}}^{\dagger }+m_{\text{in}}\right) /\sqrt{2}$, and $Y_{\text{in}%
}=i\left( m_{\text{in}}^{\dagger }-m_{\text{in}}\right) /\sqrt{2}$. The
superscript `T' denotes the transpose. The drift matrix $\mathbf{U}$ takes
the form%
\begin{equation}
\mathbf{U=}\left(
\begin{array}{cccc}
-\kappa & \Pi _{+} & 0 & J \\
\Pi _{-} & -\kappa & -J & 0 \\
0 & J & \Theta _{+} & \Xi _{+} \\
-J & 0 & \Xi _{-} & \Theta _{-}%
\end{array}%
\right) ,  \label{9}
\end{equation}%
with $\Pi _{\pm }=\pm \tilde{\Delta}_{a}-G$, $\Theta _{\pm }=-\gamma \pm K$Im%
$\left[ M^{2}\right] $ and $\Xi _{\pm }=\pm \tilde{\Delta}_{m}-K$Re$\left[
M^{2}\right] $. For a given solution of $M$ in Eq. (\ref{6}), only if all
eigenvalues of the matrix $\mathbf{U}$ have negative real parts, the
solution is said to be stable, otherwise this solution is unstable \cite{65}%
. After numerically carrying out this stability analysis [see Fig. \ref{fig2}
and related discussions], we find that the nontrivial solution $\left\vert
M\right\vert _{-}^{2}$ is always unstable in the whole parameter space,
while the solutions $\left\vert M\right\vert _{0}^{2}$ and $\left\vert
M\right\vert _{+}^{2}$ are stable in some parameter space, which suggests
that we only need to apply the solutions $\left\vert M\right\vert _{0}^{2}$
and $\left\vert M\right\vert _{+}^{2}$ to investigate the steady-state
quantum phase transition. In addition, to research the variances of magnon
mode quadratures, we define the time-dependent covariance matrix $\mathbf{V}%
\left( t\right) $ with $V_{\text{ij}}\left( t\right) =\left\langle
f_{i}\left( t\right) f_{j}\left( t^{\prime }\right) +f_{j}\left( t^{\prime
}\right) f_{i}\left( t\right) \right\rangle /2$ $\left( i,j=1,2,3,4\right) $%
. From Eq. (\ref{8a}), one can easily find the solutions of $V\left( \infty
\right) $ by solving the so-called Lyapunov equation $\mathbf{UV}+\mathbf{VU}%
^{T}=-\mathbf{D}$ with $\mathbf{D}$\ being the diffusion matrix, defined as $%
D_{\text{ij}}\left( t-t^{\prime }\right) =\left\langle \delta O_{\text{in}%
,i}\left( t\right) \delta O_{\text{in},j}\left( t^{\prime }\right) +\delta
O_{\text{in},j}\left( t^{\prime }\right) \delta O_{\text{in},i}\left(
t\right) \right\rangle /2$. The diagonal elements of the matrix $\mathbf{V}$
associate with the variance of quadratures. In our model, the variance of
magnon mode quadrature is given by $\langle \delta m^{\dagger }\delta
m\rangle $=$\left[ \left( V_{33}+V_{44}\right) -1\right] /2$ \cite{18}.

\section{Nonreciprocal quantum phase transition}

\begin{figure}[h]
\centering\includegraphics[width=8cm]{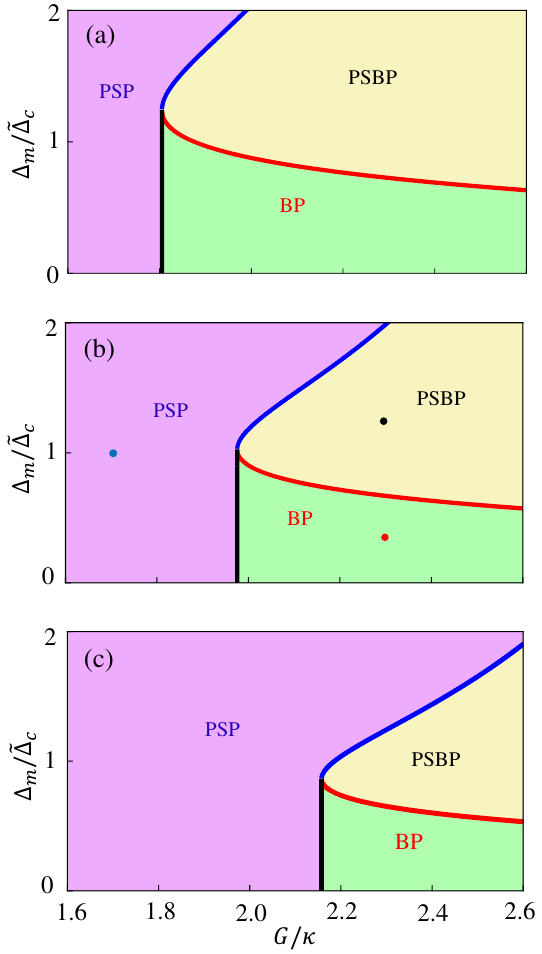}
\caption{Steady-state phase diagram of the system varying with the reduced
driving strength $G/\protect\kappa $ and the detuning ratio $\Delta _{m}/%
\tilde{\Delta}_{c}$, in which PSP, PSBP, and BP represent the areas for the
parity-symmetric phase, parity-symmetry-broken phase and bistable phase,
respectively. The Fizeau shifts are $\Delta _{F}/\protect\kappa =-0.3$ in
(a), $\Delta _{F}/\protect\kappa =0$ in (b), and $\Delta _{F}/\protect\kappa %
=0.3$ in (c). The other parameters are $\Delta _{c}/\protect\kappa =3$, $J/%
\protect\kappa =2.5$, and $\protect\gamma /\protect\kappa =1.$ }
\label{fig2}
\end{figure}

\begin{figure}[h]
\centering\includegraphics[width=8cm]{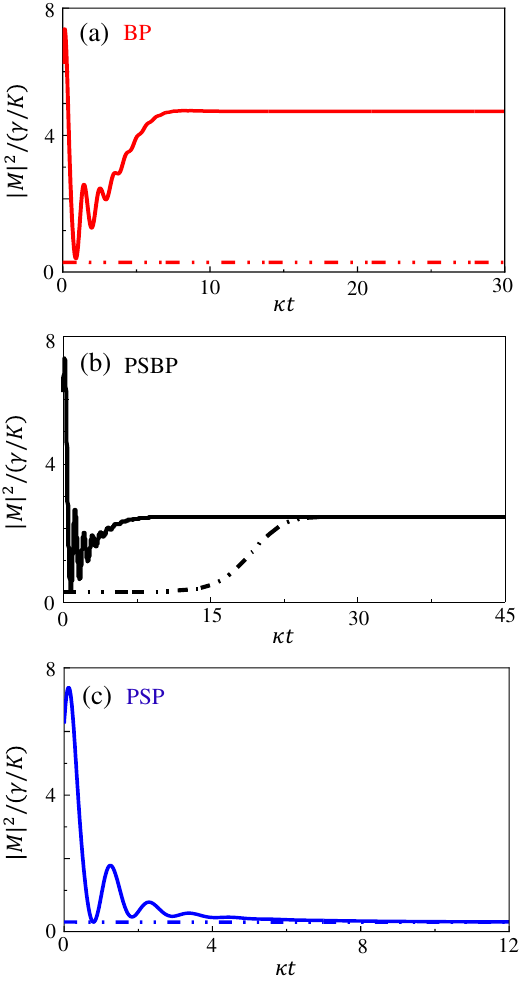}
\caption{Dynamics of the normalized magnon number for three phases at the
points marked by different-colored dots in Fig. 2(b), with (a) $G/\protect%
\kappa =1.7$ and $\Delta _{m}/\Delta _{c}=1$, (b) $G/\protect\kappa =2.3$
and $\Delta _{m}/\Delta _{c}=1.2$, (c) $G/\protect\kappa =2.3$ and $\Delta
_{m}/\Delta _{c}=0.4$. In (a)-(c), $\left\langle a\right\rangle _{t=0}/%
\protect\sqrt{\protect\gamma /K}=2+1.5i$ and $\left\langle b\right\rangle
_{t=0}/\protect\sqrt{\protect\gamma /K}=2-1.5i$ for the solid curves, while $%
\left\langle a\right\rangle _{t=0}/\protect\sqrt{\protect\gamma /K}%
=0.05+0.05i$ and $\left\langle b\right\rangle _{t=0}/\protect\sqrt{\protect%
\gamma /K}=0.05-0.05i$ for the dashed curves. The other parameters are $J/%
\protect\kappa =2.5$ and $\protect\gamma /\protect\kappa =1.$}
\label{fig3}
\end{figure}

In this section, we explore how to realize a nonreciprocal QPT via the
spinning of the resonator. We begin by employing the standard stability
analysis to show the phase diagrams against the reduced driving strength $%
G/\kappa $ and the ratio $\Delta _{m}/\tilde{\Delta}_{c}$. As illustrated in
Figs. \ref{fig2}(a)-\ref{fig2}(c), each steady-state phase diagram exists
three different regions corresponding to different phases, i.e., PSP, PSBP,
and BP.
It is worth noting that the closed system (i.e., excluding system dissipations)
without drive (for example, standard Dicke model~\cite{2}) can experience an
equilibrium QPT, whose occurrence is often accompanied by a mutation of the
ground state of the system. Different from the equilibrium QPT, the nonequilibrium
QPT occurs in the driven-dissipative system,
heralding the alteration of the steady state (rather than the ground state) of the system.
In this work, the cavity-magnon system is driven-dissipative, and each point in the phase
diagrams corresponds to stable states of the system [cf. Figs. \ref{fig2}(a)-\ref{fig2}(c)].
In the PSP (PSBP), the system is in the steady state without (with) macroscopic magnon
excitation. However, the system has two steady states in the BP, one without macroscopic
magnon excitation and the other with macroscopic magnon excitation (cf. Fig. \ref{fig3}
and related discussions). In Fig.~\ref{fig2}, the
vertical black solid line ($G=G_{\text{c1}}$) is the boundary between
PSP and BP in the case of $\Delta _{m}/\tilde{\Delta}_{a}\leq \beta |_{G=G_{%
\text{c1}}}$, and then the blue and red solid curves ($G=G_{\text{c2}}$) are
the boundaries between PSP and PSBP for $\Delta _{m}/\tilde{\Delta}_{a}\geq
\beta |_{G=G_{\text{c1}}}$, BP and PSBP for $\Delta _{m}/\tilde{\Delta}%
_{a}\leq \beta |_{G=G_{\text{c1}}}$, respectively. These manifest that the
boundaries between different phases are determined by the critical driving
strengths $G_{\text{c1}}$ and $G_{\text{c2}}$. In particular, the
intersection point of three boundaries emerges when $G_{\text{c1}}=G_{\text{%
c2}}$ at $\Delta _{m}/\tilde{\Delta}_{a}=\beta |_{G=G_{\text{c1}}}$. More
interestingly, we find that different spinning directions (the left or
right) give rise to QPT at different critical driving strength. Compared
with the stationary case (i.e., no spinning with $\Delta _{F}=0$), the areas
of three phases have a remarkable change for $\Delta _{F}>0$ and $\Delta
_{F}<0$. Therefore, this QPT are highly controllable and can be tuned by the
driving direction (or the spinning direction) and the spinning speed of the
resonator. As a result, the nonreciprocal QPT can be achieved in this system.

In Figs. \ref{fig3}(a)-\ref{fig3}(c),\ we further display the dynamical
behaviors of the scaled mean magnon number $\left\vert M\right\vert
^{2}/\left( \gamma /K\right) $ to reveal the macroscopic magnon excitations
for different phases. Here the parameter values for each phase have marked
with different-colored dots in Fig. \ref{fig2}(b). When the system is in BP
[see Fig. \ref{fig3}(a)], we find that both the solutions $\left\vert
M\right\vert _{0}^{2}$ and $\left\vert M\right\vert _{+}^{2}$ are stable and
the steady-state magnon occupation can be dominated by the initial condition
of the system.
When the system stabilizes to the steady state $\left\vert M\right\vert _{0}^{2}$ $(=0)$,
the steady state has the even parity, i.e., the parity symmetry of the system is conserved.
Conversely, since $\left\vert M\right\vert _{+}^{2}$ can be odd or even (corresponding
to the odd or even parity), it is difficult to say that the parity of the steady state is
odd or even in this case of $\left\vert M\right\vert _{+}^{2}$.
For the steady state with $\left\vert M\right\vert _{+}^{2}$, the parity symmetry is broken.
It is noticed that PSP and BP
cannot be distinguished by only tracking the dynamics of the system with an
initial state near zero since the solution $\left\vert M\right\vert _{0}^{2}$
is stable for both phases. In this case, the initial state should be set far
away from zero to distinguish the two phases and observe the phase
transition. From Fig. \ref{fig3}(b), we see that only the nontrivial
solution $\left\vert M\right\vert _{+}^{2}$ is stable in PSBP,
so that the macroscopic magnon excitation
(i.e., $\left\vert M\right\vert _{+}^{2}$) can continuously
varies and the parity symmetry of the system is broken in this situation.
As described in Fig. \ref{fig3}(c), there is no macroscopic magnon
excitation in PSP, that is to say, only the solution $\left\vert
M\right\vert _{0}^{2}$ $(=0)$ is stable and then the system is parity-symmetric.
Besides, an interesting
feature is that the nontrivial solution $\left\vert M\right\vert _{-}^{2}$
is always unstable in the whole parameter range since the four eigenvalues
of the matrix $\mathbf{U}$ in Eq. (\ref{9}) have at least one nonnegative
real part.

\begin{figure}[h]
\centering\includegraphics[width=8cm]{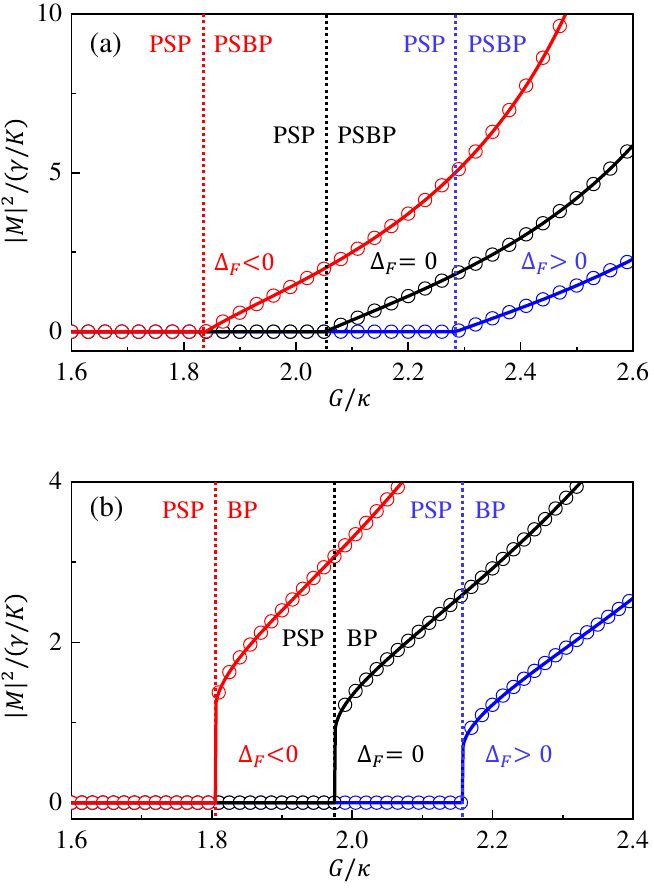}
\caption{Normalized steady-state magnon number $\left\vert M\right\vert
^{2}/\left( \protect\gamma /K\right) $ is plotted as the scaled driving
strength $G/\protect\kappa $ for three different Fizeau shifts $\Delta _{F}/%
\protect\kappa =-0.3$, $0$, and $0.3$. The solid curves correspond to the
analytical results in Eq. (\protect\ref{6d}) and the circles correspond to
the numerical results obtained by Eq. (\protect\ref{6}). (a) $\Delta _{m}/%
\protect\kappa =4$ and initial conditions $\left\langle a\right\rangle
_{t=0}/\protect\sqrt{\protect\gamma /K}=\left\langle b\right\rangle _{t=0}/%
\protect\sqrt{\protect\gamma /K}=0.2+0.2i$, (b) $\Delta _{m}/\protect\kappa %
=2.2$ and initial conditions $\left\langle a\right\rangle _{t=0}/\protect%
\sqrt{\protect\gamma /K}=\left\langle b\right\rangle _{t=0}/\protect\sqrt{%
\protect\gamma /K}=10+10i$. The other parameters are $\Delta _{c}/\protect%
\kappa =3$, $J/\protect\kappa =2.5$, and $\protect\gamma /\protect\kappa =1.$
}
\label{fig4}
\end{figure}

To exhibit the order of QPT and see its nonreciprocity more clearly, we
focus on the behavior of the order parameter $\left\vert M\right\vert ^{2}$
especially near the critical threshold for different Fizeau shifts $\Delta
_{F}$. In Fig. \ref{fig4}(a), the scaled steady-state magnon number $%
\left\vert M\right\vert ^{2}/\left( \gamma /K\right) $ is plotted as a
function of the driving strength $G/\kappa $ in the case of $\Delta _{m}/%
\tilde{\Delta}_{a}\geqslant \beta |_{G=G_{\text{c1}}}$. We find that the
critical driving strength is $G=G_{\text{c2}}$ and $\left\vert M\right\vert
^{2}/\left( \gamma /K\right) $ changes continuously from zero to non-zero,
which means that the system undergoes a second-order phase transition
from the PSP (with conserved parity symmetry) to PSBP
(with broken parity symmetry). In this process, the parity symmetry is spontaneously broken.
When $G<G_{\text{c2}}$, the system is in the PSP with $%
\left\vert M\right\vert ^{2}/\left( \gamma /K\right) =0$. However, in the
case of $G>G_{\text{c2}}$, it is in the PSBP with $\left\vert M\right\vert
^{2}/\left( \gamma /K\right) >0$. Figure \ref{fig4}(b) plots the scaled
steady-state magnon number $\left\vert M\right\vert ^{2}/\left( \gamma
/K\right) $ versus the driving strength $G/\kappa $ for $\Delta _{m}/\tilde{%
\Delta}_{a}\leqslant \beta |_{G=G_{\text{c1}}}$. We see that $\left\vert
M\right\vert ^{2}/\left( \gamma /K\right) $ has an obvious jumping behavior
across the critical point $G_{\text{c1}}$, confirming a first-order phase
transition from PSP to BP. Furthermore, it is found from Fig. \ref{fig4}
that different driving directions (the left or right) will lead to the QPT
occurring at the different critical driving strength. In comparison with the
stationary case with $\Delta _{F}=0$, the critical driving strengths of
phase transition\ of the spinning system always increases for $\Delta _{F}>0$%
, while they decreases for\ $\Delta _{F}<0$. Therefore, we can tune
(increase or decrease) the critical driving strength effectively by
adjusting the spinning direction and the spinning speed of the resonator.
More specifically, we clearly see from Fig. \ref{fig4} that the second- and
first-order phase transitions respectively occur at $G/\kappa =2.05$ and $%
1.97$ in the case of the stationary-resonator $\left( \text{i.e., }\Delta
_{F}=0\right) $. By rotating the resonator, the position of the QPTs move
towards the left (right) with $\Delta _{F}<0$ $\left( \Delta _{F}>0\right) $%
, namely, the mean magnon numbers $\left\vert M\right\vert ^{2}/\left(
\gamma /K\right) $ are respectively greater than zero from $G/\kappa >1.84$
in Fig. \ref{fig4}(a) and $G/\kappa >1.81$ in Fig. \ref{fig4}(b) when $%
\Delta _{F}/\kappa =-0.3$. $\left\vert M\right\vert ^{2}/\left( \gamma
/K\right) >0$ begin with $G/\kappa >2.28$ in Fig. \ref{fig4}(a) and $%
G/\kappa >2.16$ in Fig. \ref{fig4}(b) with $\Delta _{F}/\kappa =0.3$.
Furthermore, to confirm the validity of the analytical calculation, we also
numerically simulate the QPT behaviors of the system (the circles), which
high agrees with the analytical results (the solid curves). As a
consequence, the magnon number $\left\vert M\right\vert ^{2}/\left( \gamma
/K\right) >0$ can be achieved for driving the resonator from one direction
and $\left\vert M\right\vert ^{2}/\left( \gamma /K\right) =0$ for driving
from the opposite direction. Thus, the spinning-induced direction-dependent
magnon number can be attributed to a nonreciprocal QPT.

\

\begin{figure}[h]
\centering\includegraphics[width=8cm]{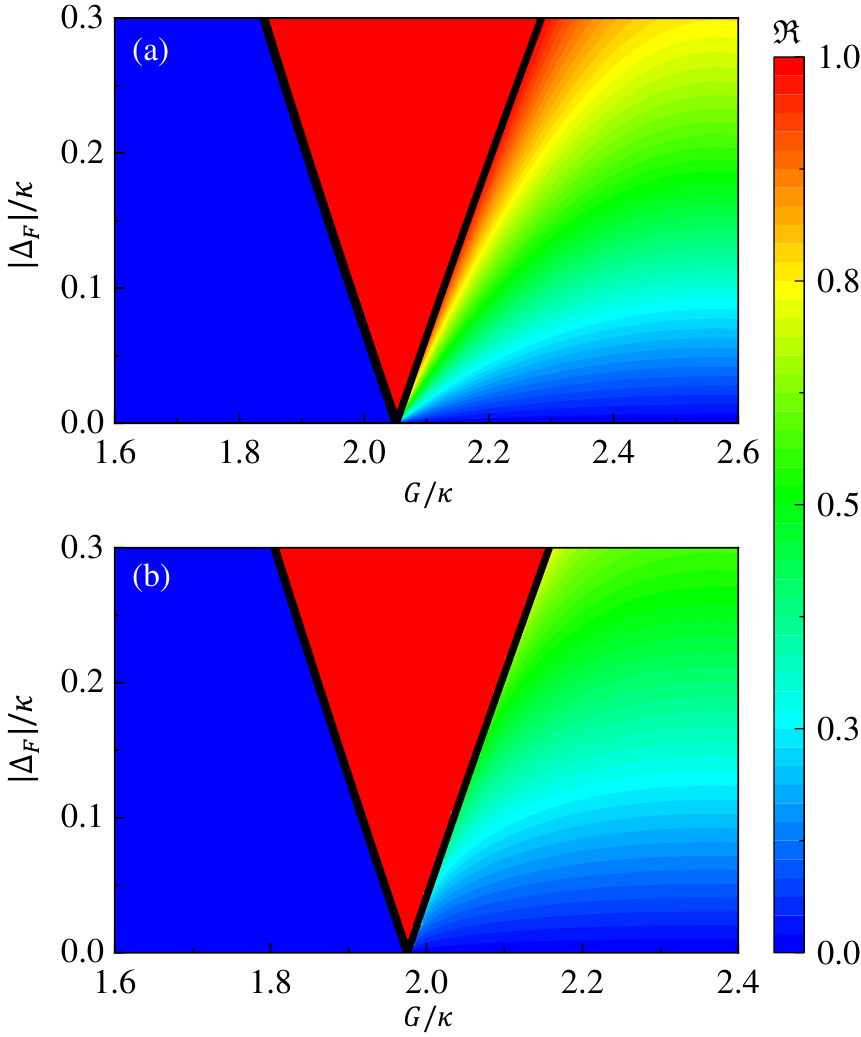}
\caption{Isolation parameter $\mathfrak{R}$ versus the scaled Fizeau shift $%
\left\vert \Delta _{F}\right\vert /\protect\kappa $ and normalized driving
strength $G/\protect\kappa $. The left and right black curves correspond to
the boundary conditions in Eqs. (\protect\ref{6d1})\ and (\protect\ref{6e}),
respectively. The parameters are the same as those in Fig. 2.}
\label{fig5}
\end{figure}

Next, to quantitatively describe the nonreciprocal QPT, we introduce the
isolation parameter \cite{47a}%
\begin{equation}
\mathfrak{R}=\left\{
\begin{array}{cc}
0,\text{ } & \left\vert M\right\vert ^{2}\left( \Delta _{F}>0\right) {\small %
=}\left\vert M\right\vert ^{2}\left( \Delta _{F}<0\right) , \\
\left\vert \frac{\left\vert M\right\vert ^{2}\left( \Delta _{F}<0\right)
-\left\vert M\right\vert ^{2}\left( \Delta _{F}>0\right) }{\left\vert
M\right\vert ^{2}\left( \Delta _{F}<0\right) +\left\vert M\right\vert
^{2}\left( \Delta _{F}>0\right) }\right\vert , & \left\vert M\right\vert
^{2}\left( \Delta _{F}>0\right) {\small \neq }\left\vert M\right\vert
^{2}\left( \Delta _{F}<0\right) .%
\end{array}%
\right.
\end{equation}%
For a QPT without the spinning resonator (i.e., a conventional reciprocal
QPT), the isolation parameter is $\mathfrak{R}=0$. A nonzero $\mathfrak{R}$
denotes the emergence of nonreciprocity in the phase transition. The higher
the isolation parameter $\mathfrak{R}$ is, the stronger the nonreciprocity
of the QPT is. Especially, $\mathfrak{R}=1$ corresponds to an ideal
nonreciprocal QPT. The isolation parameter $\mathfrak{R}$ as the function of
the Fizeau shift\ $\left\vert \Delta _{F}\right\vert /\kappa $ is shown in
Figs. \ref{fig5}(a) and \ref{fig5}(b) for the second-order and first-order
phase transitions, respectively. As expected, the isolation parameter is $%
\mathfrak{R}=0$ for the QPT with a stationary resonator (i.e., $\Delta
_{F}=0 $). While it is clear that, not only the nonreciprocal QPT occurs in
a remarkably broad parameter range, but also the region of the ideal
nonreciprocal QPT enlarges with the increasing of the Fizeau shift $%
\left\vert \Delta _{F}\right\vert $. Further, the ideal nonreciprocal QPT
exists obviously two boundaries, one boundary condition (the left black
curve) satisfies $G=G_{\text{c2}}\left( \Delta _{F}<0\right) $ and the other
one (the right black curve) satisfies $G=G_{\text{c2}}\left( \Delta
_{F}>0\right) $ in Fig. \ref{fig5}(a). In the same way, there exists a
similar characteristic in Fig. \ref{fig5}(b), where the boundary condition
(the left black curve) obeys $G=G_{\text{c1}}\left( \Delta _{F}<0\right) $
and the other one (the right black curve) obeys $G=G_{\text{c1}}\left(
\Delta _{F}>0\right) $. An interesting feature in Fig. \ref{fig5} is that in
the vicinity of the left black curses, the isolation parameter $\mathfrak{R}$
experiences a sudden change from zero to one. This is because both mean
magnon numbers $\left\vert M\right\vert ^{2}\left( \Delta _{F}<0\right) $
and $\left\vert M\right\vert ^{2}\left( \Delta _{F}>0\right) $ are zero when
the driving strength is less than the critical strength. However, when the
driving strength is more than the critical strength, $\left\vert
M\right\vert ^{2}\left( \Delta _{F}<0\right) $ is nonzero and $\left\vert
M\right\vert ^{2}\left( \Delta _{F}>0\right) $ is still zero. In this
situation, we see that the isolation parameter $\mathfrak{R}$ undergoes a
gradually change from one to zero. It can be well understood that $%
\left\vert M\right\vert ^{2}\left( \Delta _{F}<0\right) $ is nonzero and $%
\left\vert M\right\vert ^{2}\left( \Delta _{F}>0\right) $ is zero when the
driving strength is less than the critical strength. On the other hand, when
the driving strength is more than the critical strength, both $\left\vert
M\right\vert ^{2}\left( \Delta _{F}<0\right) $ and $\left\vert M\right\vert
^{2}\left( \Delta _{F}>0\right) $ are not only nonzero, but also gradually
increasing with the driving strength increasing.

\begin{figure}[h]
\centering\includegraphics[width=8cm]{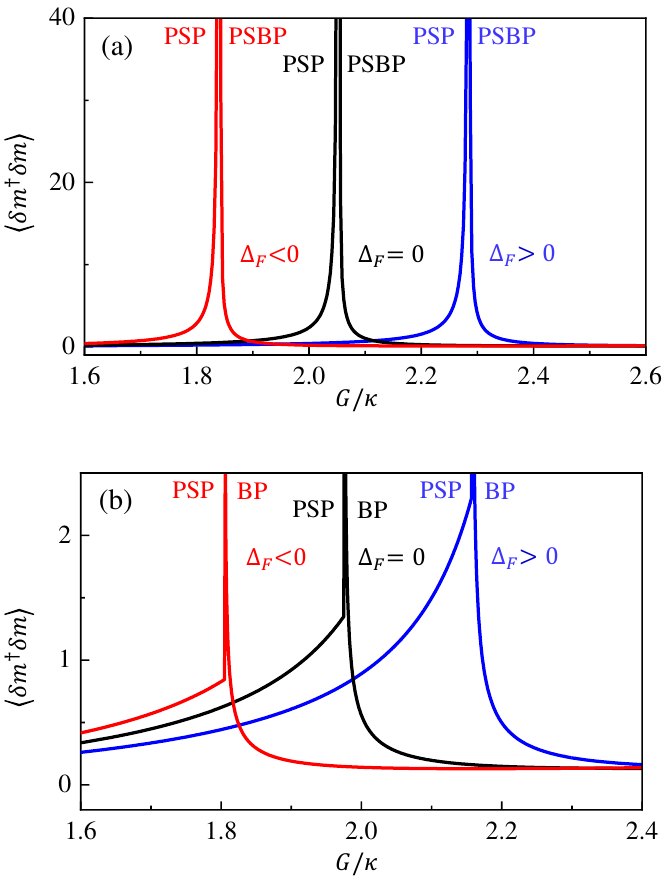}
\caption{Mean correlated fluctuation $\langle \protect\delta m^{\dagger }%
\protect\delta m\rangle $ as a function of the scaled driving strength $G/%
\protect\kappa $ for three different Fizeau shifts $\Delta _{F}/\protect%
\kappa =-0.3,0,$and $0.3$. The other parameters are the same with Fig. 4.}
\label{fig6}
\end{figure}

Finally, we reveal more characteristics of the nonreciprocal QPT through
numerically simulating the behavior of the correlated fluctuations of the
magnon operators $\delta m^{\dagger }$ and $\delta m$. As shown in Fig. \ref%
{fig6}, we plot the expectation value $\langle \delta m^{\dagger }\delta
m\rangle $ of correlated fluctuation as a function of the scaled coupling
strength $G/\kappa $ for different values of $\Delta _{F}$. There is a
common characteristic in both Figs. \ref{fig6}(a) and \ref{fig6}(b), that
is, the mean correlated fluctuation $\langle \delta m^{\dagger }\delta
m\rangle $ is close to zero when the driving strength $G$ is away from the
critical value, but Fig. \ref{fig6}(a) (Fig. \ref{fig6}(b)) is of a
asymptotic (catastrophic) divergent behavior around the critical point,
which is connected with the second-order QPT from PSP to PSBP (the
first-order QPT from PSP to BP). From Fig. \ref{fig6}, we notice especially
that spinning the resonator decreases the critical driving strength for $%
\Delta _{F}<0$ or increases it for $\Delta _{F}>0$, compared with the
stationary-resonator case ($\Delta _{F}=0$). Besides, we find from Figs. \ref%
{fig4} and \ref{fig6} that the critical positions where the QPT occurs in
Fig. \ref{fig6}(a) (Fig. \ref{fig6}(b)) completely agree with those in Fig. %
\ref{fig4}(a) (Fig. \ref{fig4}(b)) as we expected. The results give further
evidences for the nonreciprocal QPT.

\section{Discussions and conclusions}

Before concluding, we briefly discuss the experimental feasibility of the
present proposal. In cavity magnonics, the decay rate $\kappa $\
of the microwave cavity, the damping rate $\gamma $\ of the magnon mode, as
well as the coupling strength $J$\ between the photon and magnon modes are
of the order $1$~MHz (e.g., $\kappa /2\pi =2.04$\ MHz, $\gamma /2\pi =1.49$\ MHz,
and $J/2\pi =8.17$\ MHz in Ref.~\cite{67}). In
addition, the nonlinear coefficient $K$ of the magnon Kerr effect is of the order
$1$\ nHz~\cite{03a}, and the thermodynamic limit (i.e., the weak nonlinearity limit)
$\gamma /K\rightarrow \infty $\ can be easily satisfied.
By controlling the strength of the flux drive on a Josephson parametric amplifier,
the amplitude $G$ of the parametric driven can be
adjusted from $0$\ to $6$\ MHz~\cite{68}. As for the nonreciprocity induced by the Fizeau
shift, it is proportional to the spinning speed of the
resonator, for instance, a spinning speed of $12.9$\ kHz can bring about the
Fizeau shift $\Delta _{F}$\ $\approx 1.14$\ MHz \cite{69}. Therefore, our proposal
is likely to be achieved in in cavity-magnon systems with current technologies.

In summary, we have theoretically studied the nonreciprocal QPT in the
cavity magnonic system consisting of a YIG sphere placed in a spinning
microwave resonator. By rotating the microwave resonator, we show that the
introduced Sagnac effect can significantly modify the critical driving
strengths for both second- and first-order QPTs and lead to a largely
adjustable range of the critical driving strength. We further find that the
critical driving strength of the phase transition relies on both the driving
direction (or the rotation direction) and the spinning speed. In
consequence, a highly tunable nonreciprocal QPT can be achieved based on the
Sagnac effect. Our proposal provides an alternative route to achieve
nonreciprocal quantum phase in microwave magnonical system and may find
promising applications in designing nonreciprocal magnonic devices.

\section*{Acknowledgments}

This work is supported by the National Science Foundation for Distinguished
Young Scholars of the Higher Education Institutions of Anhui Province (Grant
No. 2022AH020097). G. Q. Zhang is supported by the National Natural Science
Foundation of China (Grant No. 12205069).

Note added. -In preparing our manuscript, we became aware of a similar work
on nonreciprocal superradiant phase transitions published on Physical Review
Letters \cite{61}.

\end{document}